\documentstyle[12pt]{article}
\begin{document}

\begin{center}
{\Large{\bf MESONIC AND BINDING CONTRIBUTIONS\\
\vspace{0.3cm}

 TO THE NUCLEAR DRELL-YAN 
PROCESS}}
\end{center}

\vspace{1cm}

\begin{center}
{\large{E. Marco and E. Oset}}
\end{center}

\begin{center}
{\small{\it Departamento de F\'{\i}sica Te\'orica and IFIC\\
Centro Mixto Universidad de Valencia-CSIC\\
46100 Burjassot (Valencia), Spain}}
\end{center}

\vspace{3cm}

\begin{abstract}
{\small{
We have evaluated the Drell-Yan cross section in nuclei paying special
attention to the meson  cloud contribution from pion and
$\rho$-meson, for which an accurate calculation using the meson
nuclear spectral functions is used. Similarly, the nucleonic
contribution is evaluated in terms of a relativistic nucleon spectral
function.  Fair agreement with experiment is found for different nuclei
and the results show a sizeable contribution from the renormalized
meson cloud. In order to reproduce the experiment a novel element  
is introduced, consisting of a gradual energy loss of the incoming proton
in its pass through the nucleus which produces a strong $A$ dependence
at $x_1$ large.}}
\end{abstract}

\vspace{2cm}


\newpage

The Drell-Yan process (DY) in nuclei in which a quark (antiquark) from
a proton beam and an antiquark (quark) from the constituents of the nuclear
target fuse to produce a $\mu^+ \mu^-$ pair, has been advocated as a
complement of  deep inelastic scattering (DIS) in nuclei in order
to investigate the reasons  why  the quarks in a nuclear medium
(even if they are integrated into other effective degrees of freedom,
like nucleons and mesons) behave differently than when they are in
isolated nucleons \cite{5,6,6bis,7,7bis,8,9}. 
Comprehensive reviews of this process
and its relation to the EMC effect can be seen in \cite{1,2,3,4}.

Since in the DY process one needs an antiquark for the reaction,
there was the hope that, compared to the DIS,
the nuclear DY  would be more 
sensitive to modifications of the pion cloud in the nucleus, one of the
mechanisms originally suggested to explain the EMC effect \cite{1,2,3,4}.

The calculations in ref.~\cite{6} show that the region of large $x_1$  and
small $x_2$ in the Drell-Yan process is enhanced due to medium modifications
of the pion cloud, while the region of small $x_1$  would simply
reproduce the same results as the EMC effect. In ref. \cite{7bis} the
pionic excess leads to a sharp increase of the DY ratio for large
$x_1$ and $x_2 < 0.5$. Similarly an increase of that ratio with increasing
values of $x_1 \cdot x_2$ is predicted in \cite{6bis} due to mesonic
effects.

The experimental data, however, does not show the expected enhancement
\cite{10} and recent papers try to justify the small role of the pion
cloud in this process and consequently in the EMC effect \cite{11}. On
the other hand, the mesonic contribution in the EMC effect was revived 
recently in \cite{12}, where a new evaluation was carried out 
in terms of the pion spectral function, avoiding the static approximation
(use of pion excess number) and demanding that the pion propagator
satisfies rigorous analytical properties.

The mesonic effects of ref.~\cite{12} 
showed up as an enhancement of the ratio
$F_{2A}/F_{2D}$ in the region $0.1 < x < 0.5$, in qualitative agreement
with previous findings. However, the  contribution from the renormalized
pion cloud was found smaller
than claimed in the past \cite{2,3}. At the same,
time the renormalized $\rho$-meson cloud in the medium was shown
to be as relevant as the one of the pion, and even more important in
some kinematical regions. Together with the use of an accurate nucleon 
spectral function, including relativistic effects, the analysis of 
\cite{12} showed a good agreement with experiment in the EMC region for
a wide range of nuclei.

In the present work we would like to extend the ideas of 
ref.~\cite{12} to the Drell-Yan process and establish the role of the meson
cloud in that reaction. The results show an important role of the meson
cloud in the DY process but, unlike in earlier works based upon the meson
cloud, we can now find 
agreement with  experiment.
In ref.~\cite{9} the effects of the pion cloud in the DY process are also
reanalyzed demanding similar requirements as those of ref.~\cite{12}. 
The effects of the $\rho$-meson cloud are not considered there.

 There is another important novelty in our work with respect to former
works in the DY process. Since the protons of the beam are strongly
interacting particles, they will have strong collisions with the 
nucleons much more often than they will eventually produce the 
$\mu^+ \mu^-$ pair. We shall take this into consideration here. 
We shall work outside
the shadowing region for  the variable $ x_2$ in
order to avoid the typical coherent phenomena of this
region. Outside this shadowing region the different parts of the
nuclear volume will contribute incoherently to the cross section. The
incoming proton collides strongly with the nucleons, but the proton
keeps travelling and there is no loss of flux.
However, in any of these
collisions the proton of the beam will lose a certain amount of
energy.  This will
change the variable $x_1$ and consequently the contribution to the
cross section. Let us see how this is implemented:

The cross section for $p N \rightarrow \mu^+ \mu^-  X$ is given by

$$
d^2 \sigma (p N \rightarrow \mu^+ \mu^-  X) = 
\frac{4 \pi \alpha^2 K}{9 q^2} \sum_a e_a^2 [q_a (x_1) \bar{q}_a (x_2)
$$

\begin{equation}
+ \bar{q}_a (x_1) q_a (x_2)] \, dx_1 dx_2
\end{equation}

\noindent
where the sum is over the flavours of the quarks and $x_1, x_2$ refer to
the beam and target nucleons  respectively. The variables $x_1, x_2$
indicate the fraction of the momentum carried by the quark (antiquark)
of the beam and target nucleons which fuse to create the 
virtual photon that leads to the $\mu^+ \mu^-$ pair of
momentum $q$. In an invariant form we have 

\begin{equation}
x_1 = 2 \frac{q \cdot p_2}{s} \quad ; \quad 
x_2 = 2 \frac{q \cdot p_1}{s} \quad ; \quad s \, x_1 x_2 = q^2
\end{equation}

\noindent
where $p_1, p_2$ are the fourmomenta of the beam and target nucleons
and $s = (p_1 + p_2)^2$. In the frame where the target nucleon is at
rest we can write:

\begin{equation}
x_1 = \frac{q^0}{E_1} \quad ; \quad x_2 = \frac{q^0 - q^3}{M}
\end{equation}

\noindent
where $M$ is the nucleon mass and $E_1$ the energy of the incoming nucleon.
The axis 3 is chosen along the direction of the beam.

On the other hand we shall have a distribution of $p^0_2,
p_2^3$ in the nucleus given by the nuclear spectral function.
This will give rise to a variable $x_{2N}$ which 
in the Bjorken limit can be written in terms of the static $x_2$ 
variable as

\begin{equation}
x_{2N} = 2 \frac{q  \cdot p_1}{s_N} = x_2 \frac{M}{p_2^0 - p_2^3}
\end{equation}

\noindent
which is the same relationship of $x_N$ to $x$ in
DIS \cite{12}.

The amount of energy that the nucleon of the beam loses in one
collision with the nucleons in the nucleus is difficult to
quantize. On the one hand some energy is transferred to the target
and many new particles can be created at the high energies which we are
discussing, $E_1 \simeq 800$ GeV. On the other hand some times the
nucleon can get excited to some resonant state 
in the strong collision and the resonance
may behave like a nucleon with respect to the electromagnetic 
$\mu^+ \mu^-$ process. Furthermore, 
given the scale of time in which the process occurs, the $\mu^+ 
\mu^-$ production could take place before the asymptotic final state
from the strong collision materializes. All this tells about the
difficulties in determining the equivalent energy loss for one collision.
For this reason we do not attempt to evaluate this magnitude, but use
the same DY experiment to fix it. We assume that in each 
collision, occurring 
with a probability $\sigma_{NN} \rho dl$, we lose a fraction $\beta$
of the energy. This fraction does not need to be the same for all energies,
but we will assume it to be constant in the energy range where we move.
We take the same function for computations in different nuclei. The value
of $\beta$ is taken such as to reproduce one experimental point in one
nucleus, which will be described later on.
Thus we have:

\begin{equation}
\begin{array}{l}
\frac{d E_1}{dl} = - \sigma_{NN} \rho \beta E_1\\[2ex]
E_1 (\vec{r}) = E_{1 \, in} \, \exp [- \beta \sigma_{NN} \int_{- \infty}^z
\rho (\vec{b} , z') dz']\\[2ex]
x_1 (\vec{r}) = x_1 \, \exp [\beta \sigma_{NN} \int_{-\infty}^z \rho
(\vec{b}, z') d z']
\end{array}
\end{equation}

\noindent
where $\sigma_{NN}$ is the $NN$ total cross section,
$\sigma_{NN} = 40$ mb, $\rho (\vec{r})$
the nuclear density and $\vec{b}$ the impact parameter. We have assumed
that the nucleon loses energy but keeps moving in the forward
direction, in the spirit of the eikonal approximation.
In proton elastic collisions with the nucleon 
($\sigma_{el} \simeq \sigma_{Tot}/6$) the amount of
energy lost is negligible because the cross section is very forward
peaked \cite{12a}. So, strictly speaking we should use 
$\sigma_{in}$ instead of $\sigma_{NN}$,
but since what matters is the product $\beta \cdot \sigma$
and $\beta$ is fitted to the data, we can keep using the
formalism of eq.~(5).

With this prescription and the formalism of ref.~\cite{12}
we obtain

$$
\frac{d^2 \sigma^{(N)} (p A \rightarrow \mu^+ \mu^- X)}{d x_1 dx_2}
= \frac{4 \pi \alpha^2 K}{9 q^2} 4 \int d^3 r 
\sum_a e_a^2 [q_a (x_1 (\vec{r}))
$$
$$
\times \int \frac{d^3 p}{(2 \pi)^3} \frac{M}{E (\vec{p})}
\int_{- \infty}^\mu d p^0 S_h (p^0, p) \bar{q}_a (x_{2N})
+ \bar{q}_a (x_1 (\vec{r})) 
$$
\begin{equation}
\times \int \frac{d^3 p}{(2 \pi)^3} \frac{M}{E
(\vec{p})} \int_{ - \infty}^\mu d p^0 S_h (p^0, p) 
q_a (x_{2N})] \theta (x_{2N}) \theta (1 - x_{2N})\,
\theta (1 - x_1 (\vec{r}))
\end{equation}

\noindent
where $S_h (p^0, p)$ is the hole spectral function for the nucleon
in the nucleus.

For the pion cloud contribution we have, following ref.~\cite{12}

$$
\frac{d^2 \sigma^{(\pi)} (pA \rightarrow \mu^+ \mu^-  X)}{d x_1 d x_2}
= \frac{4 \pi \alpha^2 K}{9 q^2} (-6) \int d^3 r
$$
$$
\sum_a e_a^2 [q_a (x_1 (\vec{r})) \int \frac{d^4 p}{(2 \pi)^4}
\theta (p^0) \delta Im D_\pi (q) 2 M \bar{q}_a (x_{2 \pi})
$$
$$
+ \bar{q}_a (x_1 (\vec{r})) \int \frac{d^4 p}{(2 \pi)^4} \theta (p^0)
\delta \, Im \, D_\pi (q) 2 M q_a (x_{2\pi})]
$$
\begin{equation}
\theta (x_{2 \pi} - x_2) \; \theta (1 - x_{2 \pi}) \; 
\theta (1 - x_1 (\vec{r}))
\end{equation}

The $\rho$ contribution is obtained from eq.~(7) changing $D_{\pi} (q)
\rightarrow D_\rho (q), \, x_{2 \pi} \rightarrow x_{2 \rho}$ and the 
factor ($-$ 6) by ($-$ 12).

The magnitude $\delta D (p)$ is given, following ref.~\cite{12}, by

\begin{equation}
\delta D (p) = D (p) - D_0 (p) - 
\left. 
\frac{\partial D}{\partial \rho}\right|_{\rho = 0} \cdot \rho
\end{equation}

\noindent
with $D (p)$ standing for the $\pi$ or $\rho$ propagator in the
medium and $D_0 (p)$ the corresponding free
one. Furthermore

\begin{equation}
x_{2 \pi} = x_2 \frac{M}{p_2^3 - p_2^0}
\end{equation}

\noindent
 $p_2$ referring now to the pion momenta.
The expression of $x_{2 \pi}$ compared to the one of $x_{2 N}$ in eq.~(4) has 
an apparent minus sign which comes because
the variable $p$ for pions has opposite direction as for nucleons for
convenience. The variable $x_{2 \rho}$ is given by eq.~(9) but now $p_2$ 
stands for the $\rho$ momentum. The extra factor 2 appearing for the
$\rho$ meson contribution in the cross section, with respect to the
expression for pions, accounts for the two transverse polarizations
of the $\rho$ which couple to the nucleons in our approach \cite{12}.

In eq.~(6) the $q , \bar{q}$ with arguments $x_1 (\vec{r})$ refer to
the proton beam and $q_a (x_{2N})$ refer to the nucleon target
(average over proton and neutron).

In eq.~(7), instead, the  $q , \bar{q}$ with
argument $x_{2 \pi} \, (x_{2 \rho})$ refer to the quark distributions
of the pion ($\rho$ meson, this latter one is taken the same as for
pions \cite{13,14}). The quark distributions of nucleons and pions are taken
from refs.~\cite{15} and \cite{16} respectively. For the meson propagator
and nucleon spectral function we use the same input as in the study
of the EMC effect in ref.~\cite{12}.

A small technical change is made here. The form factor accompanying the
coupling of the mesons to the $ph$, $\Delta h$ components is taken here
of the monopole type as in \cite{12}, with the same values of $\Lambda$,
but we take it static (dependent on three momentum only) to avoid
unrealistic contributions from regions in the integration where one
is close to the poles of the form factor. We should note that these form
factors are constructed in the study of the $NN$ interaction where
$q^2<0$ \cite{Holinde}. While this does not practically modify the
results in DIS, it produces some reduction of the results here. 
Particularly, the meson contribution to the DY in one nucleon, which
can be obtained by substituting $\delta D (p)$ by the last term in
eq.~(8), is more strongly reduced and reaches reasonable numbers.

Next we show some results. We have evaluated the results for
$R = 2 d \sigma_A/ A d \sigma_{D}$ corresponding to the experiment of
\cite{10}. We have evaluated the results in two different ways. First
we take a value of $x_2$ and evaluate  a weighed cross section
multiplying the cross sections at different values of $x_1$ by
the experimental acceptances \cite{17bis}. As indicated in \cite{10},
we include in the average only the regions 4 GeV $\leq \sqrt{Q^2}
\leq$ 9 GeV and $\sqrt{Q^2} \geq$ 11 GeV, in order to avoid
the regions of quarkonium resonances. Another calculation is done by
assuming $x_1 - x_2 = 0.26$ (the average value in \cite{10}) in
order to approximately extrapolate the results to the unmeasured region.
We do not evaluate $d \sigma$ for the deuteron in our local density approach
and hence we divide by the average of the cross section on the proton
and the neutron. For values of $x_2 < 0.6$ this has errors of less
than 2 $\%$ as found in the study of DIS \cite{17,18}.

We fix the fraction of energy loss  in order to 
obtain  one point of the spectrum of $^{56}$Fe,
at $x_2 = 0.15$. The resulting fraction is $\beta = 0.035$, or  3.5$\%$
energy loss. This is a small quantity but it plays a role as we
shall see.

In figs.~1,2,3 we show the results for $^{12}$C, $^{40}$Ca and $^{56}$Fe.
The experimental results are roughly reproduced
in the three nuclei. We notice the relative importance of using
the weighed cross section instead of the one using the average
value of $x_1 - x_2$, although the differences are smaller than
10$\%$. The latter cross sections show a small $A$ dependence which is
different than the one of the weighed cross sections. The ratios
obtained from the weighed cross sections are of the order of unity, as
approximately shown by the experiment, although the $^{12}$C data seems to be
larger than unity around $x_2 \simeq 0.25$. The results using the
$x_1 - x_2$ average also show a ratio bigger than unity for $^{12}$C in
this region.

It is interesting to show the effects of the meson cloud. In figs.~1,2,3 we
have separated the nucleonic contribution from the one of the pion
and $\rho$-meson clouds. As we can see, the contribution of the pion
cloud is a bit larger than the $\rho$ one.
The role of the energy loss can be seen in fig.~1. At small values of
$x_2$, taking $x_1-x_2=0.26$, it reduces the cross section in about 
10~\%. At larger values of $x_2$ the reduction is much bigger. This stronger
reduction is easily understood
since, due to the energy loss, $x_1$ increases from its original
value $x_2 + 0.26$ and eventually becomes bigger than unity where there 
is no strength for the cross section.

It is interesting to see, however what are our predictions in two
limiting cases, $x_1$ small and $x_1$ large. In fig. 4 we show the
results for the three nuclei for $x_1 = 0.01$ and $x_1 = 0.7$.
For $x_1 = 0.01$ we obtain values of $R$ very similar to those obtained
for the EMC effect, as already noticed in \cite{6}.
The $A$ dependence in this case is very weak. 
However, for $x_1 = 0.7$ the dependence of $R$ on $x_2$ is quite
different for all three nuclei. It is easy to understand this
different behaviour: for $x_1$ very small, the energy degradation
increases $x_1$ by a certain fraction but it is still very small and
this energy loss induces small changes (smaller than 10~\%).
On the contrary, when $x_1$ is close to unity, a fractional increase of
$x_1$ brings it closer to one where there is no strength. This induces
large reductions in the cross sections, which are larger in heavier
nuclei, hence the $A$ dependence seen in the cross sections.

Although our pionic effects are qualitatively similar to those in
earlier works, the energy loss changes appreciably the spectra with respect
to former predictions \cite{6,7bis}.

The issue of the energy loss in hadronic processes is attracting some
attention. Early estimates of quark energy losses based on the uncertainty
principle pointed towards very small energy losses of quarks propagating
through the nucleus \cite{20}. More accurate estimates, yet not
free of uncertainties as claimed by the authors, are done in \cite{21}.

These estimates are lower than 3.5$\%$ energy loss per
collision. More recent evaluations \cite{22} distinguish between
quarks created in the nuclear medium and incoming quarks. For the latter
case an energy loss proportional to the energy is obtained which can
be translated in our language as about 2.5$\%$ energy loss per collision.
On the other hand, the energy loss per collision counted asymptotically,
is certainly larger than 3.5$\%$ in order to reproduce the
broad energy distributions of the experimental cross sections \cite{23} (see
also comments in ref.~\cite{24} and particularly fig.~8). These 
experimental facts can be reconciled recalling our
arguments that in some cases hadronic resonances are excited, carrying
the energy of the incoming proton, which propagate through the nucleus
and can produce $\mu^+ \mu^-$ similarly as nucleons. However, in
pure hadronic reactions these resonances would decay outside
the nucleus into a nucleon and mostly pions and the nucleon
would have less energy than the incoming one.

It is also interesting  to quote that in a recent paper \cite{25} looking
 at the propagation of $J/\psi$ in nuclei a solution was favoured
implying both absorption of $J/\psi$ through the nucleus and
energy loss of the beam. This loss was equivalent to 3.6$\%$ per
collision, although some small trade-offs could be made between the 
absorption and the energy loss.

Given the relevance of the mesonic components and the energy loss in the DY 
process, we should worry about questions of selfconsistency since the
pion structure functions are determined from analysis of DY processes
in which no energy loss is assumed.
A look at fig.~4 would tell us that for values of $x_1$ small, where the
energy loss played a small role, one has not much
to worry, but for large values of $x_1$ such things could be
more relevant. This would add certain uncertainties to our
predictions for $x_1 = 0.7$ in fig.~4.

In any case the study done here clearly shows that the issue of the
energy loss is an important one and experimental efforts should be devoted
to clarify it. Fig.~4, even with accepted uncertainties, also shows
that the DY process at large values of $x_1$ is the relevant place to
look at.

In summary, our results show a sensitivity of $R$ to the meson
cloud renormalization in nuclei, more important than in the EMC effect.
The values of $R$ around unity in our interpretation are not a signal
of the lack of mesonic  effect, but the simultaneous effect of the mesonic
cloud and the progressive energy degradation of the nucleon beam
through the nucleus. The latter has the effect of reducing the cross
section since $x_1$ increases and $q (x_1) \; (\bar{q} (x_1)) $ decreases.
The present interpretation of the DY nuclear effect has as a consequence
a stronger nuclear dependence than in DIS, where the ratio is practically
constant for $A > 7$. Here we have found that for $x_1$
close to unity the ratio $R$ is rather dependent on $A$.
It would be very interesting
to have other DY experiments done with a larger range of values of
$x_1, x_2$ and better precision in order to be able to test the novel
consequences that the present interpretation of the process
provides.

\vspace{3cm}

Acknowledgments:

We would like to acknowledge useful discussions with V. Pandharipande,
G.A. Miller and S. Peign\'e. 
One of us,
E.M., wishes to acknowledge financial support from the
Ministry of Education. The work is partially supported by CICYT contract
no.~AEN 96-0753.

\newpage

\centerline{\bf{Figure Captions}}

\vspace{0.5cm}

$\bullet$ Figure 1: The ratio $2d\sigma_{A}/Ad\sigma_{D} $ for the
Drell-Yan process on $^{12}$C. Experimental points from
ref.~\cite{10}. Long-dashed line: nucleonic contribution. Dashed-dotted
line: plus pionic contribution. Solid line: plus $\rho$-meson contribution
(full calculation). The former curves use the experimental acceptances and
cuts. Short dashed line: full calculation using $x_1-x_2=0.26$.
Dotted line: full calculation with $x_1-x_2=0.26$ omitting the energy
loss.

$\bullet$ Figure 2: Same as fig.~1 for $^{40}$Ca.

$\bullet$ Figure 3: Same as fig.~1 for $^{56}$Fe.

$\bullet$ Figure 4: Three upper curves around small $x_2$: values of R for
fixed value $x_1=0.01$ for $^{12}$C, $^{40}$Ca, $^{56}$Fe (from up down),
as a function of $x_2$. The lower curves around small $x_2$: values of R
for fixed value $x_1=0.7$ for $^{12}$C, $^{40}$Ca, $^{56}$Fe (from up down),
as a function of $x_2$.

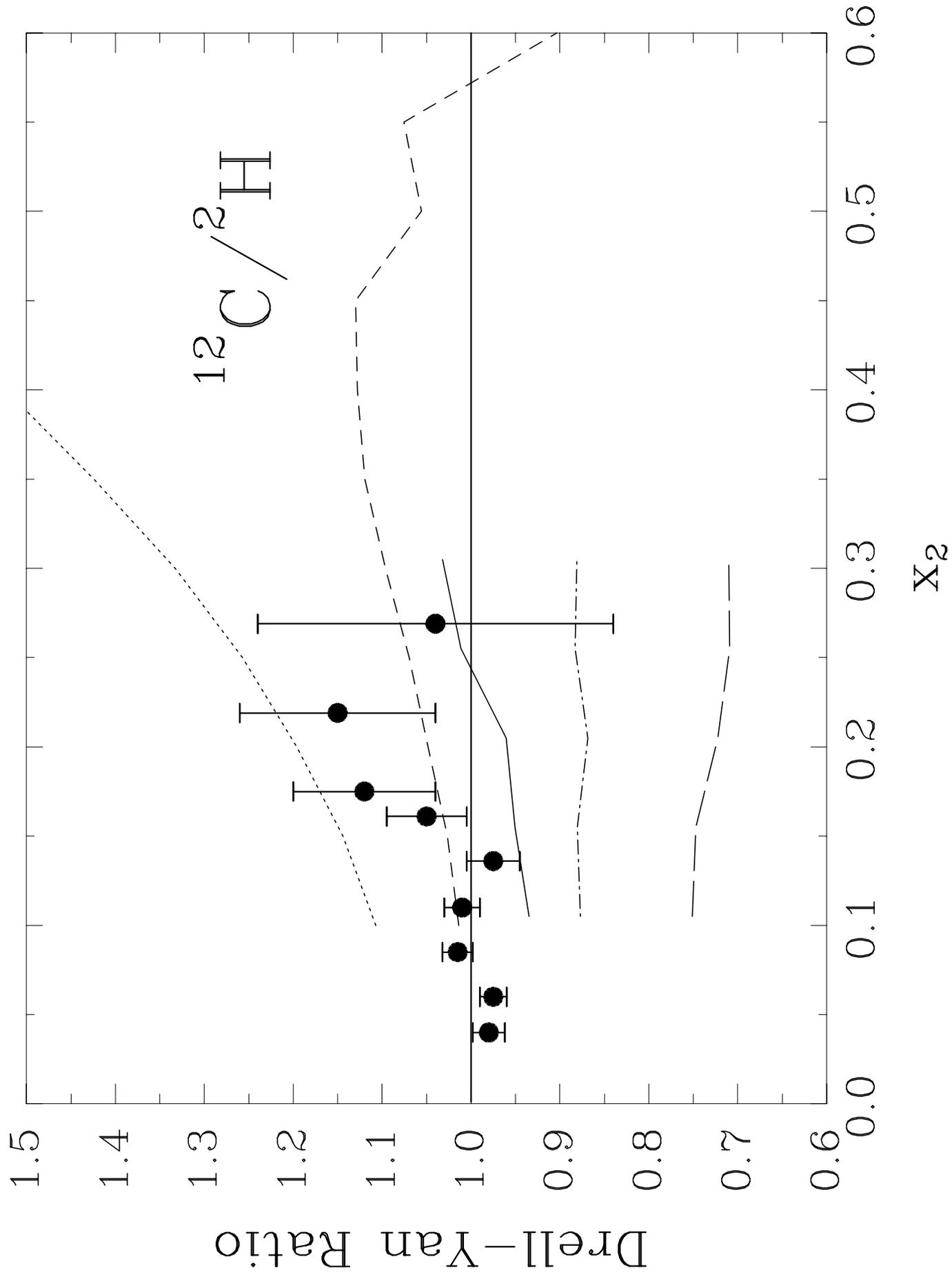
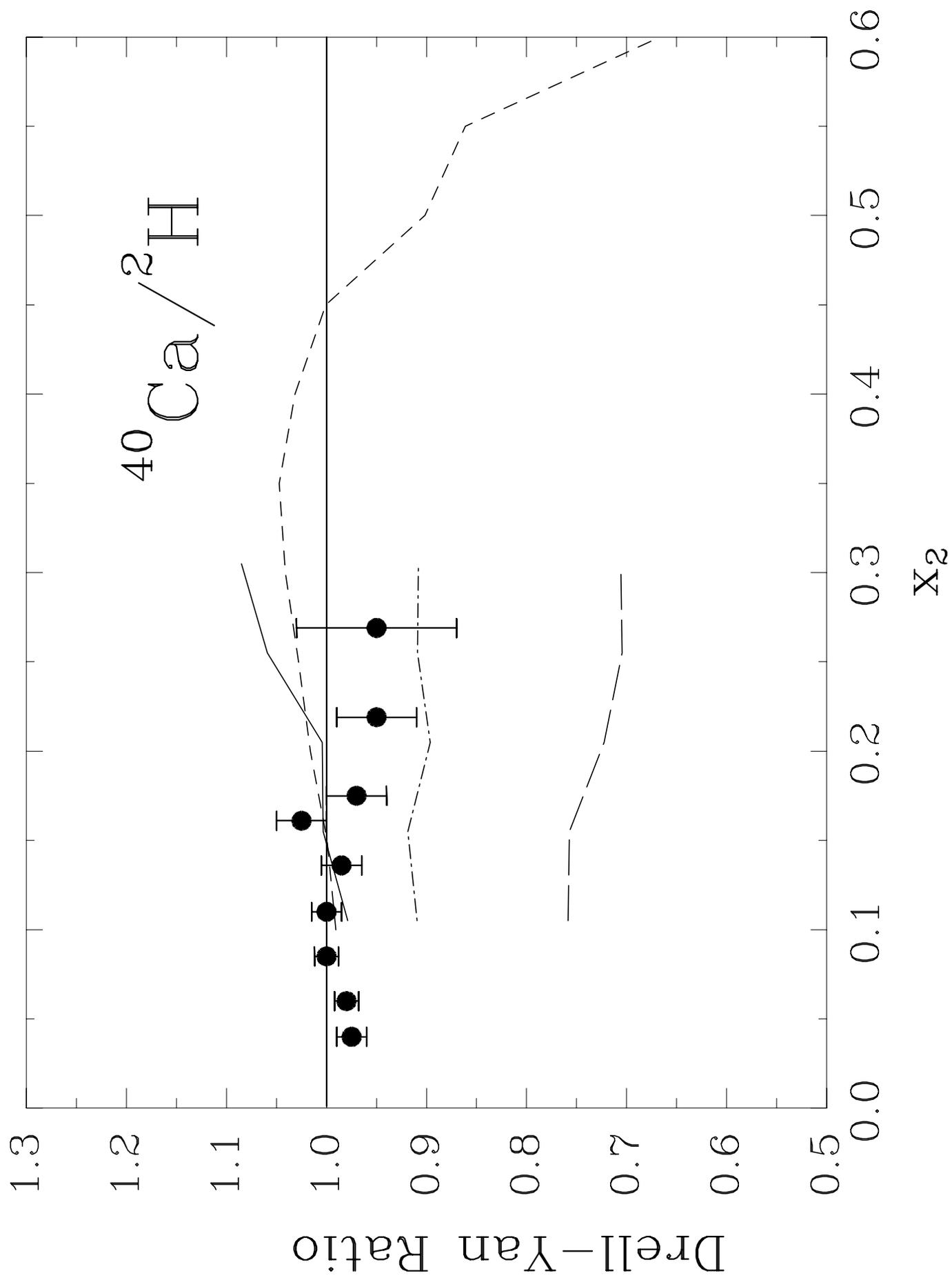
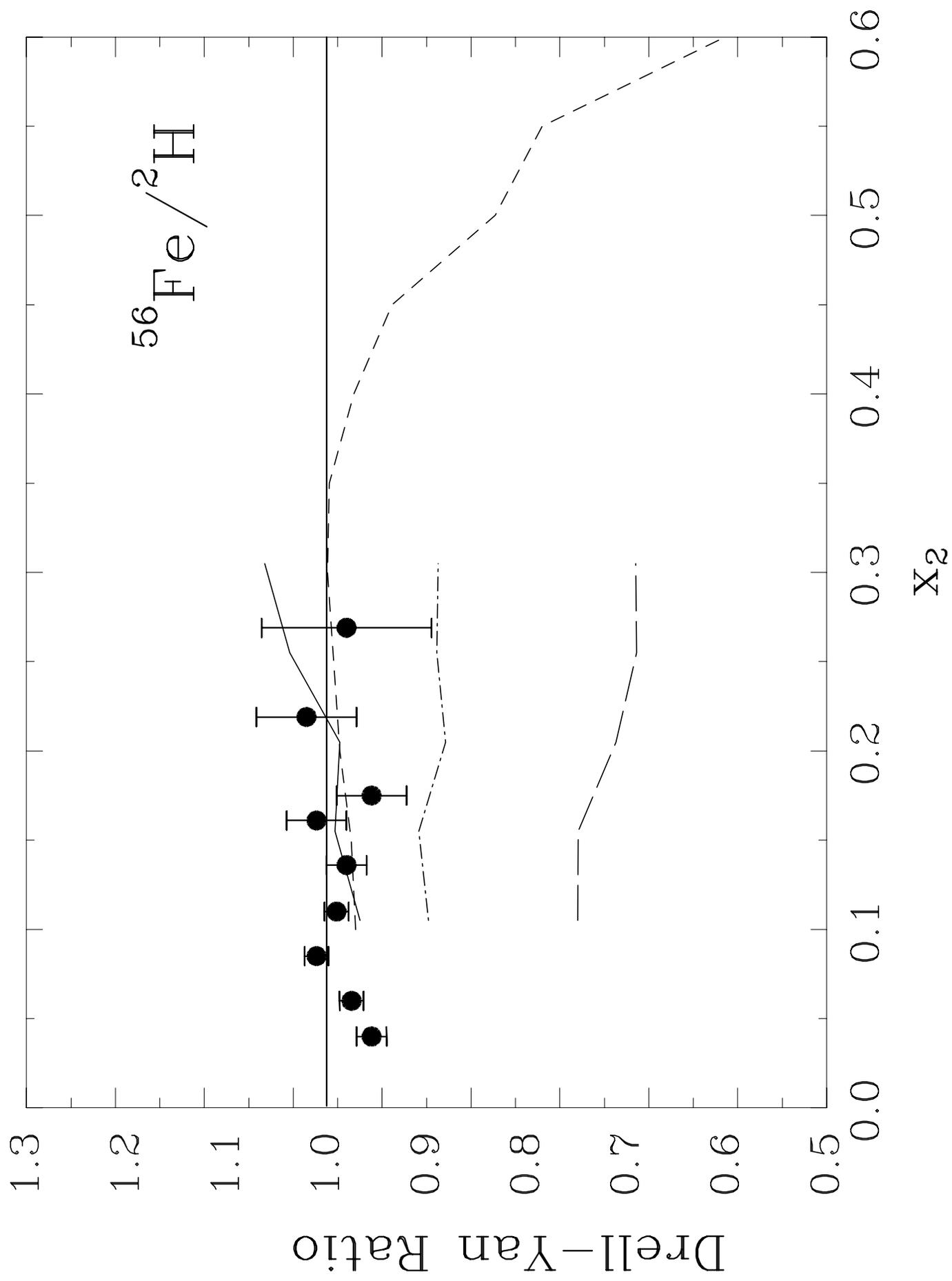
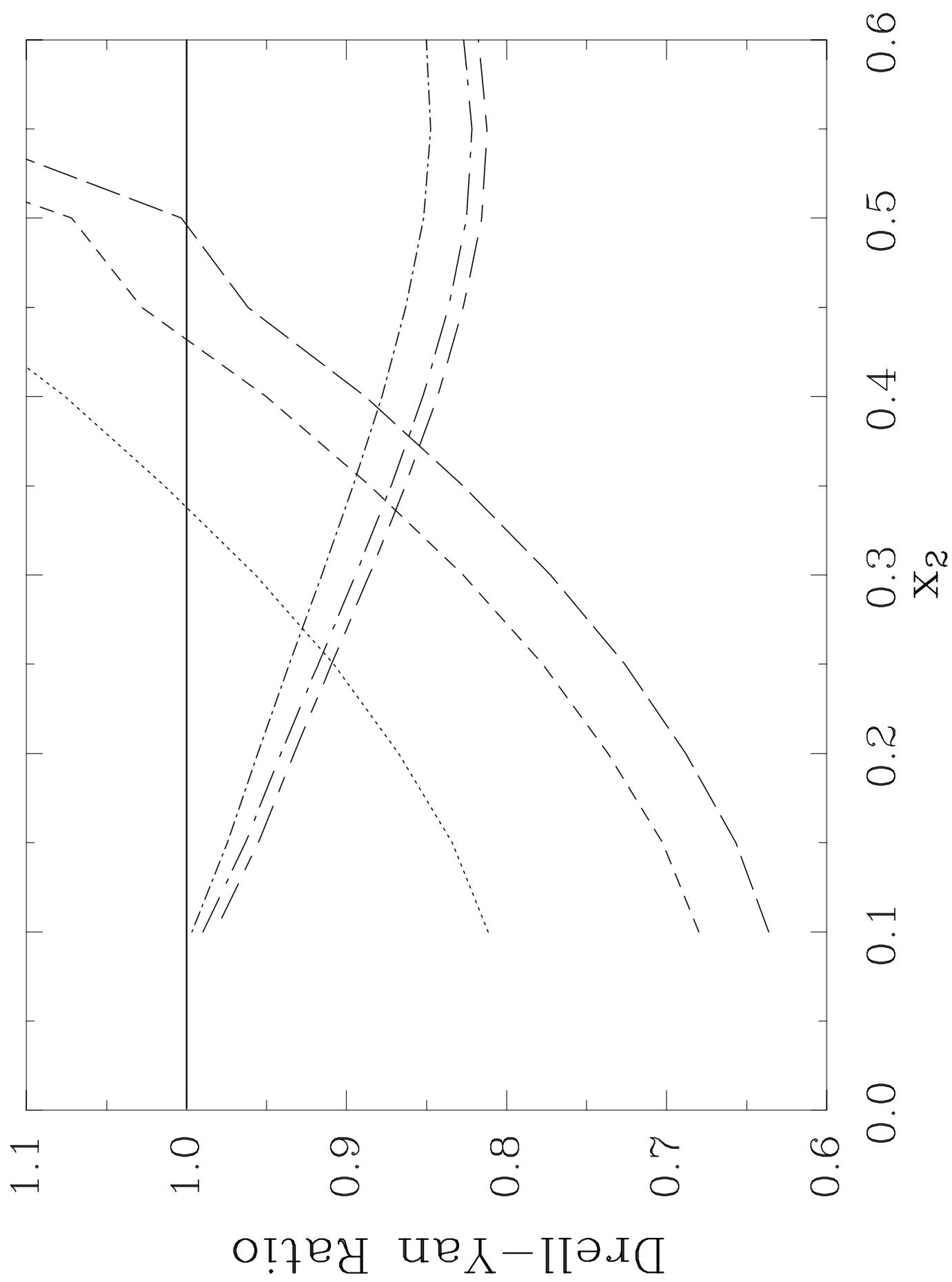
\end{document}